# Deployment of Containerized Simulations in an API-Driven Distributed Infrastructure

Making Virtual Prototypes as Accessible as Possible

Tim Kraus, Robert Bosch GmbH, Renningen, Germany (*tim.kraus@de.bosch.com*)

Axel Sauer, Robert Bosch GmbH, Renningen, Germany (*axel.sauer@de.bosch.com*)

Ingo Feldner, Robert Bosch GmbH, Renningen, Germany (*ingo.feldner@de.bosch.com*)

*Abstract*—The increasingly dynamic market for embedded systems makes virtual prototypes an indispensable tool for hardware/software codesign. The broad acceptance of the methodology has led to a diverse range of solutions: from open-source, pure console-based simulators to highly capable commercial simulation tools. In this work we present SUNRISE, an infrastructure to provide users a unified approach to utilizing virtual prototyping solutions, facilitate access to various simulation technologies and boost cooperation by leveraging decentralized compute resources for deployment of simulation workloads and definition of open APIs.

## I. INTRODUCTION

With evolving tools and methods for hardware design and with open-source IP like the RISC-V instruction set architecture [1], the semiconductor industry is undergoing an increased diversification [2]. Creating custom, application-specific semiconductors is getting more accessible and required time-to-market is getting shorter. This trend is driven further through the rapid progress in algorithmic workloads, strongly led by the ever-emerging AI algorithms, which raise the need for fitted acceleration hardware especially in highly custom sensor and edge computing products.

To enable state-of-the-art hardware/software codesign, virtual prototypes (VPs) are an essential tool throughout the development process. Depending on which development phase is being targeted, the nature of the models, that make up the virtual platforms, can vary significantly. So is the choice of simulation tool and execution environments. The categories are:
- Abstract models for early hardware design space exploration, potentially running synthetic workloads
- Functional models for software development on various abstraction levels, mostly loosely-timed (LT), running production-ready workloads without or with minor adaptions
- Accurate models in hardware (co-)verification, mostly approximately timed (AT)

Related approaches are software-in-the-loop (SiL), a method to execute embedded software on a general-purpose hardware for fast software testing and register-transfer level (RTL) simulation, which uses the exact hardware implementation to achieve the best timing accuracy. Virtual prototypes are typically developed in product specific contexts and used in certain project phases by experts who are trained on a dedicated simulation technology. A transfer of the models to another context is often initially not intended and the simulation platforms are not optimized for portability due to tool-specific intricacies. The set-up effort for users in a new context is hard to predict as it highly depends on the user's execution environment and individual competencies. In practice, it is often very time-consuming to get simulations running for new users and teams. This leads to a significant risk for the projects time schedule and less acceptance of the virtual prototyping methodology.

In this work the remote simulation framework *SUNRISE - a Scalable Unified RESTful Infrastructure for System Evaluation* - is introduced to maximize portability and simplify access to virtual development methodology without local set-up efforts. This is achieved through abstraction of the simulation details for the user by providing simple, stable and well-defined programming interfaces to configure and execute a virtual prototype in a way that is agnostic to the underlying simulation technology.

## II. RELATED WORK

A trend towards cloud-based solutions can be seen in the field of computer-aided engineering (CAE) with products like Ansys Gateway [3] or SimScale [4]. They are enabling fast and easy access to simulations with a scope towards the end user. The key elements are scalable on-demand compute resources and remote workspaces which are hidden in the back-end, enabling a flexible and convenient user experience.

Containerization and API-driven infrastructure are common techniques in the software engineering domain. However, in the digital hardware development a conventional local way of working is still the common approach. But recently, also hardware IP vendors and the EDA industry are exploring the field of modern software methods. An example is the Arm IP Explorer [5] platform, which provides a full cloud-based simulation environment for their core portfolio.

Our solution, however, is driving the open and vendor-agnostic mindset by defining open interfaces and enabling tool-agnostic integrations, whereas the above-mentioned solutions are mostly focusing on integrations inside their tool-specific ecosystems. An interaction with these solutions might require proprietary adapters or wrappers and will multiply the integration effort for each vendor, tool and IP provider.

Besides the commercial solutions, open-source frameworks for infrastructure management and workflow implementation are available. Apache Airflow [6] can be used to define generic workflows and infrastructure components. Jenkins [7] is a CI/CD management platform for generic tasks. They overlap with some features of our approach but differ in a specialization for cloud-based simulations of digital hardware. Apache Airflow and Jenkins could be used on top of SUNRISE to manage and control workflows by using its APIs.

## III. WORKFLOW

The scope of the SUNRISE approach is to enable the configuration and execution of virtual prototypes to generate result data in a straightforward and reproducible way. Therefore, the users follow the generic sequence of logical steps shown in Figure 1, which is agnostic of the simulation technology.

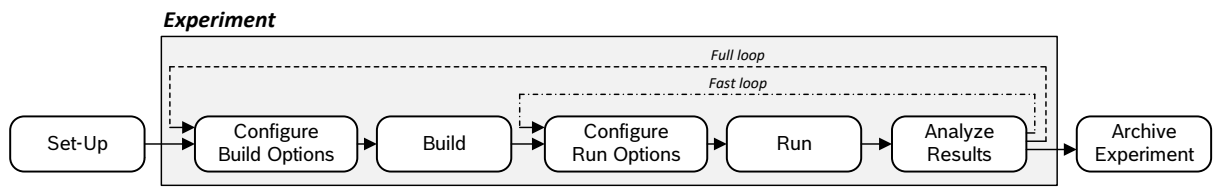

Figure 1: SUNRISE Workflow

Initially, a one-time *set-up* step is required in which the user chooses a virtual prototype from a library of available pre-assembled platforms and creates a new unique experiment with it. An experiment is a temporary session, that holds a working environment owned by a user to evaluate one use-case. The lifetime of an experiment is typically in the range of minutes to few days.

The first steps inside an experiment are the parameterization and the *build* of the simulation model. For VPs that are available in form of source code in a compiled programming language, like C++ / SystemC, this comprises the compilation and the generation of the executable. Build parameters are characterized by the fact that they have an impact on the source code, like preprocessor options, or control the behavior of the toolchain, so adjusting the parameters will require a re-compilation. For VPs that are completely pre-built and already available as executable out of the box, the build step can be dispensable.

With the simulation built and ready to be executed, there is the option to set parameters for the *run*. These parameters do not require re-building the binary, like command line arguments of the simulator. Other examples are stimulus data files which are read at run-time from the simulation or embedded software binaries that should be executed on an instruction-set simulator. With these parameters set, the VP can be executed for a defined time or until it terminates. In the SUNRISE context, the run is considered as non-interactive and automatable, so user interaction, e.g. with a graphical simulation tool, is not intended in this work.

After the run the next step involves evaluating the *results* generated by the simulator. This is highly application specific and can include log files, processor performance metrics like number of instructions and cycles, embedded software profiling, hardware signal trace files, binary data files or any kind of proprietary artifacts. Depending on the results, the user can decide to repeat the execution with modified parameters. Since build and run are separate steps, the user has the choice between fast iterations with only modifying run parameters or a longer loop that includes also the VP build.

If a final configuration has been found, the experiment can be closed and *archived*. In this way, the exact configuration under which the results were created is frozen and can be referenced at a later point in time.

## IV. INFRASTRUCTURE

The SUNRISE infrastructure is designed strongly modular to realize a scalable environment with clear separation of responsibilities. APIs are a central part of the concept, as bringing modules from different sources smoothly together requires clearly defined interfaces for all parts. As shown in Figure 2, the four sub-components are:

- One central *Runtime Manager* (RM)
- One or multiple *front-ends*
- A set of different *systems*, stored in *system storages*
- One or multiple compute *back-ends*

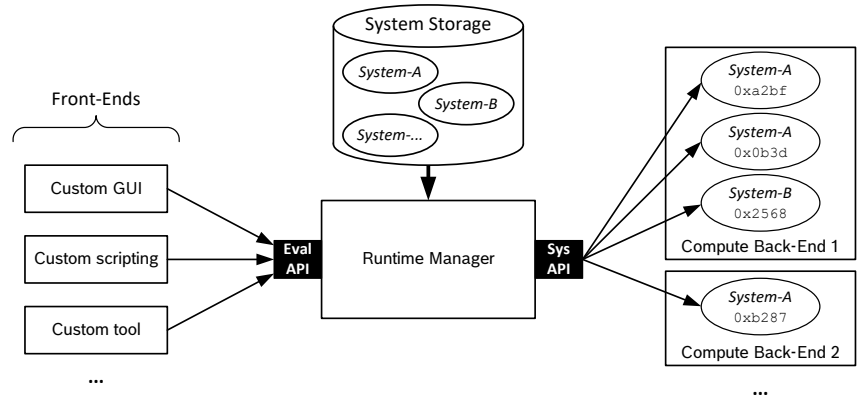

Figure 2: SUNRISE framework overview

The modular infrastructure and the interfaces are designed to enable distributed operation. This means that the components can be hosted on one single, but also on multiple separate physical or virtual machines. They can also be integrated into a container orchestration platform. An important aspect to note is, that although the overall framework can be distributed, it is not intended to split up the execution of one virtual protype between multiple compute hosts, as this would imply changes to the simulation technology with all associated challenges like timing- and data synchronization. This would contradict the goal of being independent of simulation technology, so it is not in the scope of this work.

### A. Systems

A *system* in SUNRISE is an application, typically a simulation platform with its simulation tool, and all required dependencies in form of a container image. The proof-of-concept implementation uses Docker, so it is a requirement that a simulator can be run in a non-interactive way (headless) inside a container to integrate it as a system in SUNRISE.

The properties of a system are defined in a *System Definition* JSON file, *SysDef*. This file is the foundation of the *SysAPI* which is the common interface to any system. Listing 1 shows the SysDef for an open-source virtual prototype with a programmable RISC-V core [8] integrated into SUNRISE. The schema of the JSON file is standardized, so the general handling of the file can be well automatized independent of the systems. The SysDef contains all meta information about a system, the configurable parameters and the result artifacts which are generated during the run. The instructions to build and run the system inside its container context as well as the container image itself are also required entries. As described for the workflow previously, the configurable parameters are separated between build and run. Parameters of type string, number or bool can be natively described in JSON, but for file parameters additional information is required, as it is shown for the run parameter "app" in Listing 1. For this system the embedded software is compiled to a binary file separately and uploaded to the system through this file parameter before the run. Every parameter has a default value, which can be overwritten by the user during the workflow in an experiment.

```json
{
  "name": "AGRA RISC-V",
  "version": "1.0",
  "documentation": {
    "contact": "T. Kraus",
    "summary": "RISC-V Demonstration VP",
    "description": "A platform with RISC-V core and a basic set of peripherals."
  },
  "docker_image": "some_docker_registry.com/eval_platform_agra:ready-to-run",
  "build_command": "python build.py",
  "run_command": "source run.sh",
  "build_parameters": {
    "compile_args": "-O3 -Wall"
  },
  "run_parameters": {
    "run_time_ms": 1000,
    "app": {
      "value": "demo_sw/demo_app",
      "is_file": true
    },
    "simulator_args": "--intercept-syscalls"
  },
  "results": {
    "signal_trace": {
      "path": "vp/install/sim_trace.vcd",
      "type": "vcd"
    }
  }
}
```

Listing 1: Example System Definition (SysDef) JSON file

The systems container image, which is referenced in the SysDef, is taken from a separate container registry. The SysDef file itself is stored on a file storage or in a version control system which is accessible by the Runtime Manager.

### B. Runtime Manager

The *Runtime Manager (RM)* is the central application that brings the components in the infrastructure together. It operates the *SysAPI* to execute systems on compute back-ends and it provides the *EvalAPI* towards the user front-ends. A session management is implemented to keep track of parameters used in a simulation during an experiment, as described in the workflow definition (Figure 1). For unambiguous identification, the RM assigns a Universally Unique Identifier (UUID) to each experiment. Metadata is added to an experiment, like the username of the creator, the creation time, a textual description of the purpose and the current status. To realize the whole virtual prototyping workflow, the RM implements a state machine for each experiment, from which it can derive which operations are possible: e.g. run requires a finished build and result artifacts can only be retrieved if a run has finished.

The exact implementation of the RM is subordinary and can be done in any way, as long as it operates the APIs according to their definition. For the proof-of-concept implementation of the RM, Python is chosen as programming language to make use of the numerous available libraries to implement REST-APIs, use the git version control management to pull systems, use Docker locally or access APIs of cloud services. It is a logical consequence of the concept to also containerize the RM itself, so the deployment of the overall framework is not limited to a specific host system but flexible and cloud enabled.

### C. Front-Ends

The entry point for users of virtual prototypes in the SUNRISE framework are the *front-ends*. As the use-cases strongly vary depending on project type and personal preferences of the users, these tools are highly specific and not standardized by the SUNRISE framework. A front-end can be a graphical tool that enables the user to interactively execute the virtual prototyping workflow step by step. It can also be a non-interactive custom scripting environment or a larger automation system. By defining the *EvalAPI* as a REST API to the Runtime Manager, the framework is accessible by any solution than can operate the interface.

### D. Compute Back-Ends

The execution of a system inside its container environment is done on a *compute back-end*, which accordingly must provide a container runtime engine. It is possible to use local on-premise, proprietary remote or general purpose cloud resources. To integrate a back-end, the Runtime Manager must be able to operate the specific API for the container runtime of the platform, so for a wide compatibility, the reference implementation uses a generic *compute interface* module that can be specialized for back-end types. If multiple back-ends are available, the Runtime Manager is responsible to decide which one to use. Since it is a common scenario that multiple experiments are active simultaneously and the compute interface supports on-demand compute resources, a highly scalable and cost-efficient infrastructure can be achieved. Through the SUNRISE framework, technical details on how to operate a specific back-end are hidden in the front-end for the user.

In a simple, fully local approach without remote compute back-ends, the compute interface can work on a local container engine operated e.g. over the Docker API. This way, the scalability of compute power is traded for the certainty that all confidential data that is part of the system and the experiment remains on the local machine. If the SUNRISE infrastructure itself is hosted with an orchestration engine, the compute interface can connect to this already available and technically mature platform and use it for the deployment of the simulation workloads. E.g., Kubernetes [9] is a well-established solution to abstract cloud resources and provide a unified container runtime environment.

### E. Application Programming Interfaces (APIs)

The APIs define the structure of the SUNRISE framework and are the main element to enable modularity and therefore separation of responsibilities and ownership of components.

The *System API (SysAPI)* introduces a way to access a system in SUNRISE, that is independent from the technology used inside the system container. The definition comprises the specification of meta information about a system in form of the System Definition file (Listing 1), as well as the way the system container is accessed to configure parameters, build the simulation, execute it and extract result artifacts. With this set of information, the RM is able to operate the system by calling the build or run command over the entry point of the system container. The concrete parameter values to be used in an experiment are added to a System Configuration (SysCfg) JSON file which is derived from the SysDef file. A set of options for common result types, like VCD in Listing 1, is defined to allow re-use of analysis tools. The results can be extracted over the SysAPI by defining the location of the result file inside the System Definition.

The *Evaluation API (EvalAPI)* is the exposed interface towards the framework users. The interface is provided as REST paradigm and comprises all functionality to operate the systems of the system storage with a specific parameterization. Such a web API is highly flexible and supported by many tools and libraries used by front-end developers. Table 1 shows an excerpt of the significant endpoints. Having this API is a key benefit this work introduces for virtual prototyping users, as it is independent of the simulation model used as system, the simulation technology behind and the compute back-end on which the workload is executed.

| Endpoint | Method | Parameters | Response | Description |
|---|---|---|---|---|
| **/session** | POST | SysCfg, Description (optional) | Experiment UUID | Create a new experiment for a specific system configuration (SysCfg) |
| **/session/{session_id}/parameter** | POST | Experiment UUID, Parameter name | - | Upload a file-based parameter |
| **/session/{session_id}/build** | POST | Experiment UUID, Timeout (optional) | - | Build the system |
| **/session/{session_id}/run** | POST | Experiment UUID, Timeout (optional) | - | Run the system |
| **/session/{session_id}/status** | GET | Experiment UUID | Status | Returns the status of an experiment (e.g. *running* or *completed*) |
| **/session/{session_id}/result/{name}** | GET | Experiment UUID, Result name | File object containing the result | Request a result as file |

Table 1: Excerpt of Evaluation API endpoints

## V. APPLICATION EXAMPLES

### A. System Benchmarking and Comparison

One common application of SUNRISE is the performance evaluation of processors or complete SoCs (System on Chips) with benchmark workloads on accurate simulation platforms. This is a headless use-case, since the benchmark software is typically executed as a whole to create quantitative result without user interaction.

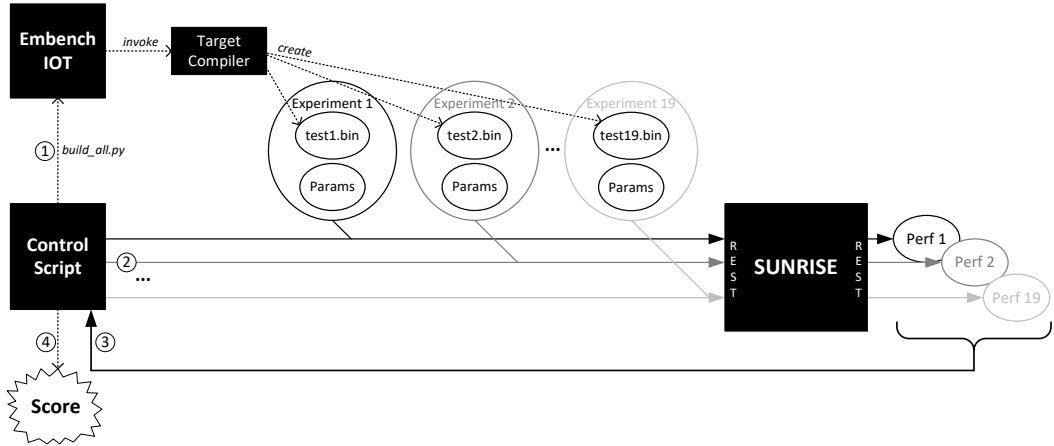

Figure 3: Parallelization of a benchmark with SUNRISE

Figure 3 shows how the execution of Embench IOT [10] can be done on the framework. A control script ① first calls the build mechanism of Embench to generate the binaries for the target hardware. This leads to 19 binaries, representing 19 test applications to be executed on the target ②. As SUNRISE can scale up the workload execution, it is possible to also start 19 experiments in parallel. After the simulation runs finished, the control script collects the performance metrics of all sub-benchmarks ③ and calculates the overall score ④.

Since the flow is completely independent of the simulation technology, which is used inside the system, the approach can be easily reused for other cores in other systems by just replacing the target compiler. SUNRISE thereby supports creating a database of performance values for a collection of cores.

## B. HW/SW Codesign for AI application

Creating a solution for data processing on custom hardware can be seen as a multi-objective, multi-parameter optimization task. As shown in Figure 4, in hardware/software codesign for an AI based application, subject for optimization is the algorithm (AI Model), the hardware details and the mapping of the workload on the hardware. The criteria to optimize for are typically performance, power and area (PPA).

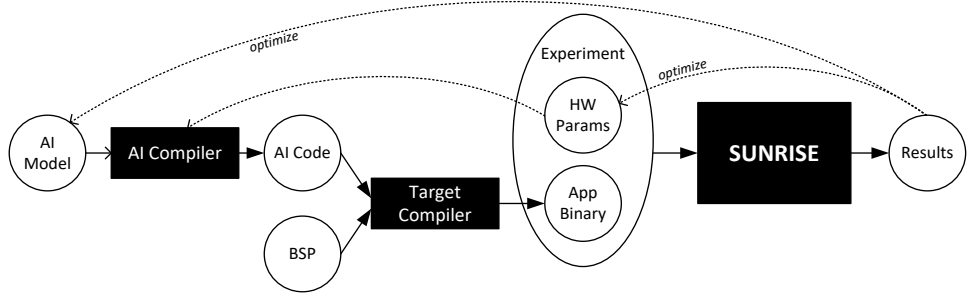

Figure 4: SUNRISE to optimize AI processing

When using SUNRISE for HW/SW codesign, the system exposes relevant hardware configuration options (HW Params) which are used in the hardware aware AI compiler. The output of the AI toolchain is fed into the target compiler together with a board support package (BSP) software to create an executable binary to run on the target system. After execution, the results are used to optimize workload algorithm and hardware parameters. The process to find an optimal solution can be automated and controlled through optimization methods of the user's choice. Through the access to hardware parameters, common neural architecture search (NAS) approaches are extended by another dimension allowing to find an overall better solution. By using a capable compute back-end like high-performance compute centers, the optimization can be massively parallelized and thus substantially accelerated. Using the EvalAPI significantly simplifies this for users from the algorithmic domain by hiding the complexity of the simulation infrastructure.

## C. Integration into DevOps Environments

In contrast to the previous example, where SUNRISE was applied in early project phases, the framework is also beneficial for development of product software by using it as part of a DevOps environment. During development, continuous testing is essential and virtual prototypes are a well-suited replacement for physical hardware prototypes, which are only available late, in limited numbers and at high cost. Although, as soon as the hardware was selected, the parameters of the virtual prototype configuration itself are not suspect to change, other system parameters that supply stimulus data for tests are well applicable in this context. Using SUNRISE to access virtual prototypes in a continuous integration / continuous deployment (CI/CD) pipeline, which is triggered by changes on a source code management (SCM) server, is advantageous.

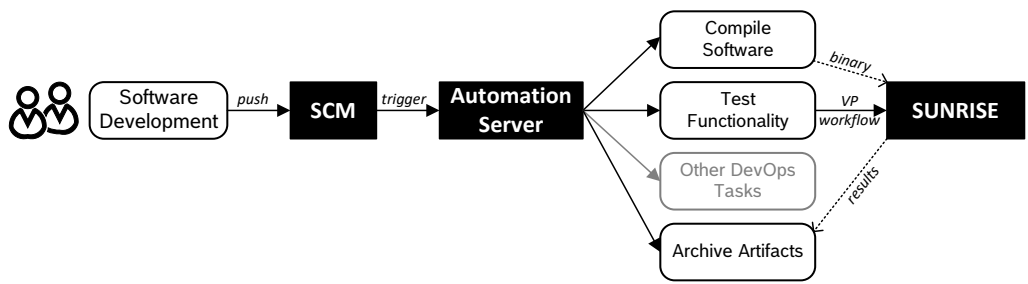

Figure 5: SUNRISE in a CI/CD pipeline

On the one hand, since the EvalAPI to access the simulation is stable and independent from the simulation model and technology, there is high level of re-use between product variants and projects, saving the CI/CD engineer from repeated integration tasks for the VPs. By developing convenient and universal plugins for SUNRISE (e.g. Jenkins Plugins or GitHub Actions), the integration effort is minimized. On the other hand, the scaling of compute resources for simulations through our framework reduces run times of large testbenches significantly, so the software developer receives the feedback about the latest code changes faster and the overall software development efficiency is improved. Although some automation servers also have similar features, SUNRISE provides a unified solution for executing simulations than can be used across different CI/CD platforms.

## VI. CONCLUSION

The decentralized execution environment has proven to provide efficient and flexible access to virtual prototyping. By specifying the infrastructure and APIs on a higher level, a common ground for numerous use-cases is covered, independent of the simulation technology that is used. On the front-end, graphical control and visualization dashboards were connected as well as custom automation scripts. The containerization of computation workloads in the back-end has proven to enable fast set-up of simulations and scaling for highly parallel execution.

An upcoming feature is the introduction of specialized gateway systems that run on dedicated back-ends which have access to physical hardware resources like FPGAs. This will require an extension of the API and back-end definition by the annotation of available specialized features.

## ACKNOWLEDGMENT

This work is partially funded by the German Federal Ministry of Education and Research (BMBF) in the project MANNHEIM-FlexKI (grant number: 01IS22086A).

## REFERENCES

[1] RISC-V International, accessed 2024-04-18, https://riscv.org/about/
[2] John L. Hennessy and David A. Patterson. 2019. A new golden age for computer architecture. Commun. ACM 62, 2 (February 2019), 48–60. https://doi.org/10.1145/3282307
[3] Ansys Gateway, accessed 2024-06-18, https://www.ansys.com/products/cloud/ansys-gateway
[4] SimScale, accessed 2024-06-18, https://www.simscale.com/
[5] Arm IP Explorer, accessed 2024-06-18, https://www.arm.com/products/ip-explorer
[6] Apache Airflow, accessed 2024-06-14, https://airflow.apache.org
[7] Jenkins, accessed 2024-06-14, https://www.jenkins.io/
[8] Vladimir Herdt, Daniel Große, Pascal Pieper, Rolf Drechsler. 2020. RISC-V based virtual prototype: An extensible and configurable platform for the system-level. Journal of Systems Architecture, 109, 101756. https://doi.org/10.1016/j.sysarc.2020.101756
[9] Kubernetes, accessed 2024-06-24, https://kubernetes.io/
[10] Embench IOT Repository, release embench-1.0, accessed 2024-04-18, https://github.com/embench/embench-iot